\documentclass[preprint,superscriptaddress,showpacs,preprintnumbers,amsmath,amssymb,aps,pre]{revtex4-2}
\usepackage{epsfig}
\usepackage{natbib}
\usepackage{amsmath}
\usepackage{times}
\usepackage{subfigure}
\usepackage{graphicx}
\usepackage{color}
\usepackage{dsfont}
\usepackage{lipsum}
\graphicspath{{./figures/}}

\begin{document}

\title{Relativistic chaotic scattering: unveiling scaling laws for trapped trajectories}

\author{Fernando Blesa}
\affiliation{Departamento de Física Aplicada,
University of Zaragoza, 50009 Zaragoza, Spain}

\author{Juan D. Bernal}
\affiliation{Nonlinear Dynamics, Chaos and Complex Systems Group, Departamento de
F\'{i}sica, Universidad Rey Juan Carlos, Tulip\'{a}n s/n, 28933 M\'{o}stoles, Madrid, Spain}

\author{Jes\'{u}s M. Seoane}
\email[]{jesus.seoane@urjc.es}
\affiliation{Nonlinear Dynamics, Chaos and Complex Systems Group, Departamento de
F\'{i}sica, Universidad Rey Juan Carlos, Tulip\'{a}n s/n, 28933 M\'{o}stoles, Madrid, Spain}

\author{Miguel A. F. M. Sanju\'{a}n}
\affiliation{Nonlinear Dynamics, Chaos and Complex Systems Group, Departamento de
F\'{i}sica, Universidad Rey Juan Carlos, Tulip\'{a}n s/n, 28933 M\'{o}stoles, Madrid, Spain}

\date{\today}

\begin{abstract}

In this paper, we study different types of phase space structures which appear in the context of relativistic chaotic scattering.
By using the relativistic version of the H\'{e}non-Heiles Hamiltonian, we numerically study the topology of different kind of exit basins and compare it with the case of low velocities in which
the Newtonian version of the system is valid. Specifically, we numerically study the escapes in the phase space, in the energy plane and also in the $\beta$ plane which richly characterize the dynamics of the system. In all cases, fractal structures are present, and the escaping dynamics is characterized. Besides, in every case a scaling law is numerically obtained in which the percentage of the trapped trajectories as a function of the relativistic parameter $\beta$ and the energy is obtained. Our work could be useful in the context of charged particles which eventually can be trapped in the magnetosphere, where the analysis of these structures can be relevant.

\end{abstract}

\maketitle

\section{Introduction} \label{sec:Introduction}

\textit{Chaotic scattering} in open Hamiltonian systems is a relevant topic in nonlinear dynamics and chaos, which has been broadly studied during the last decades. There have been many applications of interest in physics (see Refs.~\cite{New Developments,OTT}). In general terms, the problem of chaotic scattering is defined as the interaction between an incident particle and a region described by a potential or a massive object. The aforesaid region is usually named \textit{scattering region} and the potential or massive object, \textit{scatterer}. Oftentimes, the interaction between the incident particle and the scatterer is modelled by nonlinear equations and the resultant dynamics can be chaotic. Thus, slightly different initial conditions may describe radically different trajectories which result in diverse final conditions or destinations of the particles. The influence of the scattering region over the incident particles can be considered negligible outside this region and then the motions of the particles are uniform. Here, we are always dealing with open systems so the scattering region possesses exits from which the particles may enter and escape. From this point of view, chaotic scattering may be studied as a physical manifestation of transient chaos ~\cite{Yorke_Physica,Directions_in_Chaos}.\\

On the other hand, when the velocity of the incident particle is low compared to the speed of light, the convention in physics is the use of the Newtonian approximation to model the dynamics of the system \cite{OHANIAN}. Nonetheless, even at low velocities, when the dynamics of the system is chaotic, trajectories predicted by the Newtonian approximation may rapidly disagree with the ones described by the special relativity theory ~\cite{Lan, solitons, Borondo, bouncingball}. It is worth to note that there are relevant global properties of chaotic scattering systems such as the average escape time of the particles, their decay law, as well as the basin topology that strongly depend on the effect of the Lorentz transformations ~\cite{JD relativistic, JD relativistic topology}. Specifically, in Ref.~\cite{JD relativistic}, the authors study how the average escape times of the particles from the scattering region varies with respect to $\beta$, finding a crossover point at $\beta \simeq 0.4$ in which the KAM islands are completely destroyed. Moreover, in that work, the study of the decay law of the particles in the scattering region has two different behaviors depending of the value of $\beta$. In the case of low values of $\beta$, that decay law obeys a potential law of the form $R\sim t^{-\alpha}$, and when $\beta \geq 0.4$ this law becomes exponential in the form $R \sim e^{-\tau t}$. A mathematical relationship between $\tau$ and $\alpha$ was found as well, corroborating the numerical simulations obtained there.\\
On the other hand, a study of the Wada property of the exit basins associated to the phase space in the relativistic regime was studied in Ref.~\cite{JD relativistic topology}. In that work, the authors suggested, by computing the exit basins, the existence of Wada basins in phase space for values of $\beta < 0.625$. Furthermore, the evolution of the exit basins
by computing the basin entropy, was also studied. This last result shows that there is a maximum value of the basin entropy for $\beta \approx 0.2$ that is related to the prediction of the final state of the system.
Therefore, in case you want to describe a chaotic system in a realistic manner, the special relativity scheme have to be considered, even for low velocities. Similar conclusions have been also demonstrated when there are weak gravitational interactions among the scattering region and the particles, then the theory of general relativity should take center stage ~\cite{Kovacs 2011, JD General Relativity}. In these last papers, the authors have studied the Sitnikov model and the basins associated to its phase space, denoting both the regions of particles escaping from the influence of the gravitational force and the regions in which the particles are trapped. Besides, they found in \cite{JD General Relativity} values of the gravitational radius $\lambda$ for which a bifurcation appears. This bifurcation is related to the metamorphosis of the KAM islands for which the escape regions change when the gravitational radius takes the values of $\lambda \simeq 0.02$ and $\lambda \simeq 0.028$.\\

However, in the relativistic case, the study of the basin structures associated to other planes, such as the $(x,y)$ and $(\beta, y)$ planes, among others, has not been considered. Moreover, in this specific context, an interesting topic to analyze, in different kind of structures, is the fraction of initial conditions trapped for a suitable range of both parameter values and initial conditions of physical interest. In particular, the significance of this investigation lies in exploring regions encompassing both initial conditions and parameter values around the scattering region. This holds true for appropriate energy values in both the Newtonian and the relativistic regimes, making the study highly valuable. This is relevant since it provides us useful information on how many trajectories are escaping or not in that specific structure. Furthermore, the finding of new scaling laws, of a different nature as the ones found in Ref.~\cite{JD relativistic}, reveals how the set of initial conditions and parameter values for which the particles are trapped in the scattering region, are very relevant for a better understanding of relativistic chaotic scattering problems. This can offer novel and valuable insights that have not been analyzed in previous investigations.

Therefore, we add here the study of different fractal structures showing relevant information on both the dynamics and the topology of the system. Specifically, we study, besides the well-known $(x,y)$ and $(y, \dot{y})$ planes, the basins in the energy plane, $(E,y)$, and in the $\beta$ plane, $(\beta, y)$, where $\beta$ is the relativistic parameter defined as $\beta \equiv v/c$. In our context, we will consider two kinds of energies: $E_{N}$ is the energy of the Newtonian system ({\it Newtonian energy}) and $H$ is the energy of the relativistic system ({\it relativistic energy}), which will be properly defined in the next section. We also define the quantity $F$ as the fraction of the remaining particles in the scattering region with respect to the total number of chosen particles as a function of time. On the other hand, the study of these aforementioned structures show important insights on the evolution of the system. The different uncovered scaling laws providing important relationships between the fraction of the trapped particles $F$ and the energies $E_N$ and $H$ and the relativistic parameter $\beta$ constitute our main findings. The analysis of these structures can be useful in the study of trajectories of charged particles trapped in the Earth's magnetic field \cite{OZTURK:2012}. A reference work on fractal structures in nonlinear dynamics with applications to numerous physical systems can be seen in Ref.~\cite{RMP:2009}. Notice that, throughout this paper, we refer as \textit{relativistic} to any situation where the Lorentz transformations have been considered. Likewise, we say that any property or object is \textit{Newtonian} when we do not take into consideration the Lorentz transformations but the Galilean ones.

This paper is organized as follows. In Sec.~\ref{sec:Model Description}, we describe  the relativistic H\'{e}non-Heiles system that is the model used in our research work. Section~\ref{section_3} presents the exit basins in the phase space and the physical space characterizing the regions in which the particles are trapped. The role of the energy on the escaping dynamics is developed in Sec.~\ref{section_4}. The study of the escapes in the plane $(\beta, y)$ is carried out in Sec.~\ref{section_5} in which we clearly see the effects of high velocities on the escaping dynamics. In all previous situations a scaling law characterizing both the escape and the trapping regions are obtained. Finally, a discussion and the main conclusions of this manuscript are presented in Sec.~\ref{sec_conclusions}.


\section{Model description} \label{sec:Model Description}

Now, we describe a prototypical model to study the effects of the relativistic corrections in chaotic scattering phenomena, the well-known H\'{e}non-Heiles system ~\cite{Henon_Heiles}. It is a two-dimensional Hamiltonian system whose potential is defined by

\begin{equation}\label{eq:HH potential_1}
V(x,y)= \frac{1}{2}k(x^2+y^2) + \lambda (x^2y - \frac{1}{3}y^3).
\end{equation}

Here, we take the version of this model in which we consider a unit mass particle, $m=1$, and $k=\lambda=1$. Therefore, the system of units is arbitrary and the potential can be written as follows:

\begin{equation}\label{eq:HH potential}
V(x,y)= \frac{1}{2}(x^2+y^2) + x^2y - \frac{1}{3}y^3.
\end{equation}

\begin{figure}[h]
\centering
\includegraphics[width=0.65\textwidth]{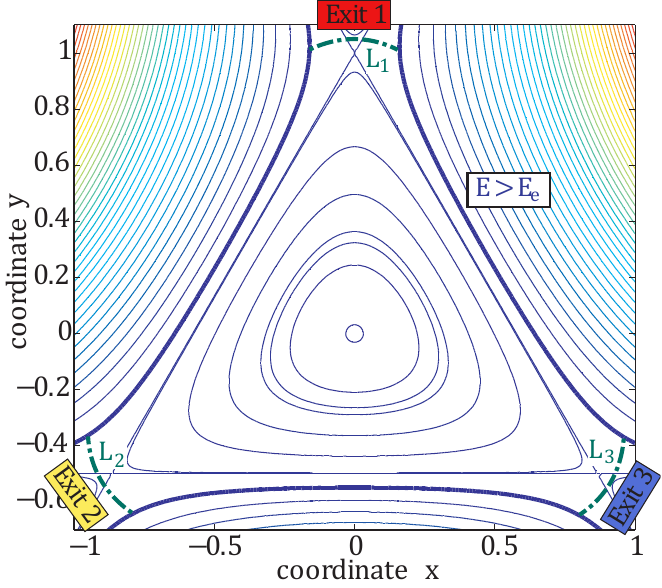}
\caption{\textbf{Isopotential curves for the H\'{e}non-Heiles potential:} they are closed for energies below the Newtonian threshold energy escape $E_e=1/6$. It shows three different exits for energy values above $E_e=1/6$. The dotted lines are the unstable Lyapunov orbits $L_i$.}
\label{Fig1}
\end{figure}

The isopotential curves of the H\'{e}non-Heiles potential can be seen in Fig.~\ref{Fig1}. For values of the energy above the critical energy $E_e=1/6$, there are three exits through which the particles may go in and out. They are separated by an angle of $2\pi/3$ radians due to the triangular symmetry of the system. We call Exit $1$ the upper exit ($y \rightarrow +\infty$), Exit $2$, the left one ($y\rightarrow -\infty, x\rightarrow -\infty$), and, Exit $3$, the right exit ($y\rightarrow -\infty, x\rightarrow +\infty$).

Along this paper, we define the Newtonian total mechanical energy, also called it here, Newtonian energy, $E_N$. This variable is defined here as $E_N = T(\mathbf{p}) + V(\mathbf{r})$, where $T$ is the kinetic energy of the particle, $T = \mathbf{p}^2/2m$, $\mathbf{p}$ is its linear momentum, $V(\mathbf{r})$ is the potential energy and $\mathbf{r}$ is its vector position. When $E_N\in [0,1/6]$, the trajectory of any incident particle is trapped in the scattering region. For $E_N > 1/6$, the particles may eventually escape up to infinity. Thus, there are three different regimes of motion depending on the initial value of the energy: \textit{(a)} closed nonhyperbolic $E_N \in [0,1/6]$, \textit{(b)} open nonhyperbolic $E_N \in (1/6,2/9)$ and \textit{(c)} open hyperbolic $E_N \in [2/9,+\infty)$ \cite{022blesa}. In the first energy range, all the trajectories are trapped and there is no exit by which any particle may escape. When $E_N \in (1/6,2/9)$, the energy is high enough to allow escapes from the scattering region, although there coexists stable invariant tori and chaotic saddles, which typically results in an algebraic decay law in the survival probability of a particle in the scattering region ~\cite{JD relativistic}. However, when $E_N \in [2/9,+\infty)$, the regime is hyperbolic, and all the periodic trajectories are unstable; there are no KAM tori in the phase space.

The motion of a relativistic particle moving in an external potential energy \textit{V(\textbf{r})} is described by the following Hamiltonian (or the total energy we also call as relativistic energy $H$):

\begin{equation}\label{eq:Hamiltoniano relativista}
H = E = \gamma mc^2 + V(\mathbf{r}) = \sqrt{m^{2}c^{4}+c^{2}\mathbf{p}^{2}}+V(\mathbf{r}),
\end{equation}

\noindent where \textit{m} is the particle's rest mass, \textit{c} is the speed of light and $\gamma$ is the Lorentz factor which is defined as:
\begin{equation}\label{eq:Lorentz factor}
\gamma = \sqrt{1 + \frac{\mathbf{p}^2}{m^2c^2}} = \frac{1}{\sqrt{1 - \frac{\mathbf{v}^2}{c^2}}}.
\end{equation}

Then, Hamilton's canonical equations are:
\begin{equation}\label{eq:ecuaciones de Hamilton}
\begin{aligned}
\dot{\mathbf{p}} & = - \frac{\partial H}{\partial \mathbf{r}} = -\nabla V(\mathbf{r}),\\
\dot{\mathbf{r}} = \mathbf{v} & = \frac{\partial H}{\partial \mathbf{p}} = \frac{\mathbf{p}}{m\gamma},
\end{aligned}
\end{equation}

Obviously, if $\gamma = 1$, the Newtonian equations of motion are recovered from Eqs.~\ref{eq:ecuaciones de Hamilton}. Here is defined $\beta$ as the ratio $v/c$, where $v$ is the modulus of the velocity $\mathbf{v}$. Then the Lorentz factor can be rewritten as $\gamma = \frac{1}{\sqrt{1 - \beta^2}}$. Whereas $\gamma\in[1,+\infty)$, the range of values for $\beta$ is $[0,1]$. In any case, $\gamma$ and $\beta$ express essentially the same thing: how large is the velocity of the object when compared to the speed of light. Here, we use $\beta$ instead of $\gamma$ to show our results for mere convenience.

Taking into consideration Eqs.~\ref{eq:HH potential} and \ref{eq:ecuaciones de Hamilton},  the relativistic equations of motion of a scattering particle of unit rest mass $(m=1)$ interacting with the H\'{e}non-Heiles potential are:

\begin{equation} \label{eq: relativistic eq motion}
\begin{aligned}
\dot{x} & = \frac{p}{\gamma},\\
\dot{y} & = \frac{q}{\gamma},\\
\dot{p} & = -x - 2xy,\\
\dot{q} & = -y - x^2 + y^2,
\end{aligned}
\end{equation}

\noindent where \textit{p} and \textit{q} are the two components of the linear momentum \textit{\textbf{p}}.\\

\begin{figure}[htp]
\centering
\includegraphics[width=0.65\textwidth,clip]{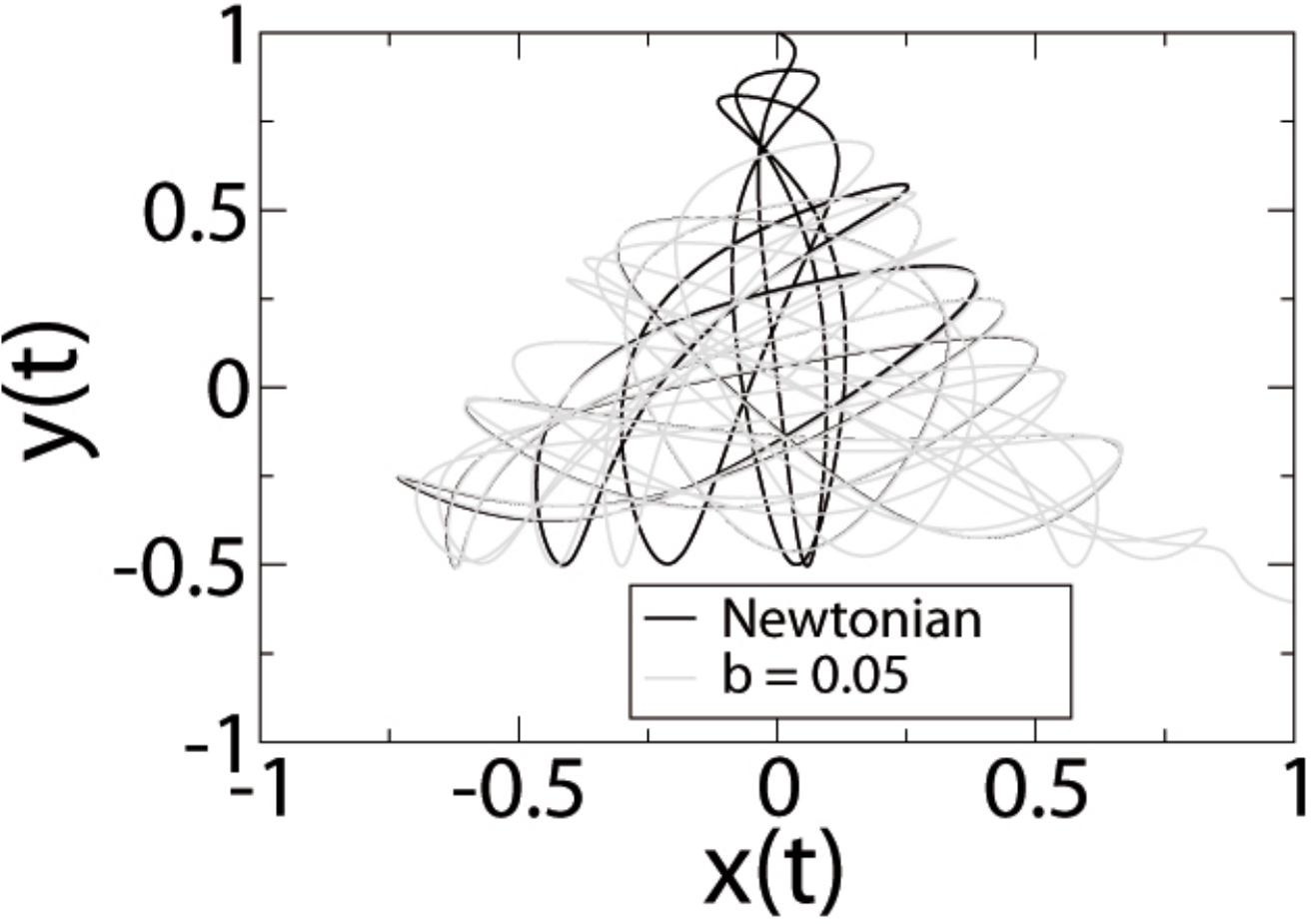}

\caption{\textbf{Comparison between two trajectories:} in black, we show the relativistic trajectory and, in pale gray, the Newtonian one. Trajectories corresponding to particles shot from the same initial condition $(x_0,y_0,\dot{x_0},\dot{y_0})= (0,0,v\cos(0.3\pi),v\sin(0.3\pi))$ with velocity $v = 0.583$. The effect of $\gamma$ in Eq.~\ref{eq: relativistic eq motion}, even for a very low velocity ($\beta = 0.05$), yields a significantly different result. Despite the consideration of the Galilean corrections results in a trajectory going out of the scattering region by Exit 1 (see Fig.~\ref{Fig1}), the same initial condition contemplating the relativistic corrections yields a trajectory that leaves the scattering region after a finite time by Exit 2.} \label{Fig2}
\end{figure}

In the present work, we aim to isolate the effects of the variation of the Lorentz factor $\gamma$ (or $\beta$ as previously shown) from the rest of variables of the system, \textit{i.e.} the initial velocity of the particles, its energy, etc. For this reason, during our numerical computations we will have to use a different system of units so that $\gamma$ is the only parameter in the equations of motion (Eq.~\ref{eq: relativistic eq motion}) that may vary. This system of units is also arbitrary where the rest energy is $E=mc^{2}=1$. Therefore, we will analyze the evolution of the properties of the system when $\beta$ varies, comparing these properties with the characteristics of the Newtonian system. For example, the initial velocity, $v = 0.583$, in different system of units, corresponds to a Newtonian energy $E_N=0.17$, which lies in the open nonhyperbolic regime and quite close to the limit value $E_e$. As an example and for the sake of clarity, we consider an incident particle coming from infinity to the scattering region.  The objective of our numerical computations and analysis is to study the effect of $\gamma$ in the equations of motion, so the key point is to set the speed of light \textit{c} as the limit value of the speed of the particles, regardless of the system of units we may be considering. In order to give a visual example, in Fig.~\ref{Fig2}(a), we represent two different trajectories of the same relativistic particle when it is shot at the same initial Newtonian velocity, $v = 0.583$, from the same initial conditions, $x = 0$, $y = 0$ and equal initial shooting angle $\phi = 0.1\pi$. However, that velocity, measured in different systems of units, represents different values of the parameter $\beta$. The black curve is the trajectory of the particle for $\beta = 0.01$ and, in light gray, we represent the trajectory for $\beta = 0.1$. Both trajectories leave the scattering regions for the same exit, although the paths are completely different and the time spent in this region too. This is aligned with the previous research ~\cite{Lan, solitons, Borondo, bouncingball}.\\

The value of $\beta$ (or $\gamma$) changes with time as the particle’s velocity $v$ also changes with time. Therefore,  it is not a constant. However, the initial value of $\beta$ is the most relevant value since it determines the initial conditions of the system and the initial deviation from the Newtonian dynamics. Furthermore, the variation of $\beta$ in time is not very large, since the potential energy $V(x,y)$ is small compared to the rest energy $mc^2$ , and the kinetic energy is bounded by the total energy $H$. Therefore, we use $\beta$ as a parameter to study the relativistic effects in the chaotic scattering dynamics.

As it has been discussed earlier in the literature, there are relevant effects of external perturbations such as noise and dissipation in the escape basins topology of some open Hamiltonian systems ~\cite{effectnoise, MOTTER:LAI}. For the sake of clarity, it is worth to note that the consideration of the relativistic framework on the system dynamics cannot be considered as an external perturbation like noise or dissipation, although the global properties of the system also change.\\


Understanding the dynamics of a physical system often requires studying its behavior in different spaces. In the following sections we describe the structures that appear in the relativistic Hénon-Heiles system when we vary different parameters. The structures are defined as the geometrical features of the exit basins and their boundaries in phase space. The exit basins are the regions of initial conditions leading to escape through a given exit, and their boundaries are fractal sets that separate different basins (specifically the boundaries are the stable manifold of the chaotic saddle \cite{New Developments, RMP:2009}). The structures reflect the complexity and unpredictability of the chaotic scattering process, and can be characterized by various quantities, such as the uncertainty dimension, the Wada property, and the basin entropy.

We have computed several planes to visualize the structures: the phase space plane ($y$, $p_y$), which shows the initial conditions in terms of the vertical coordinate $y$ and its conjugate momentum component $p_y$; the coordinate plane ($x$,$y$), which shows the initial conditions in terms of the horizontal and vertical coordinates $x$ and $y$; and the ($y$,$H$) and ($y$,$\beta$) planes, which show the initial conditions in terms of $y$ and either the relativistic Hamiltonian $H$ (we will call as relativistic energy) or the relativistic factor $\beta$. The energy $H$ measures the total energy of the system, which depends on $\beta$ and the Newtonian energy $E_N$.

In this paper, we examine how the structures change when we vary one parameter while keeping another constant. First, we examine the escapes in phase space and in the coordinate plane when the energy is kept constant. Next, we will analyze the escapes as the energy varies for certain values of $\beta$. We compare our results with those obtained for the Newtonian Hénon-Heiles system, in order to gain insights into the effects of relativity on chaotic systems.

\section{ESCAPES IN PHASE SPACE $(y,p_y)$ AND IN PHYSICAL SPACE $(x, y)$}\label{section_3}

\begin{figure}
\includegraphics[width=\textwidth]{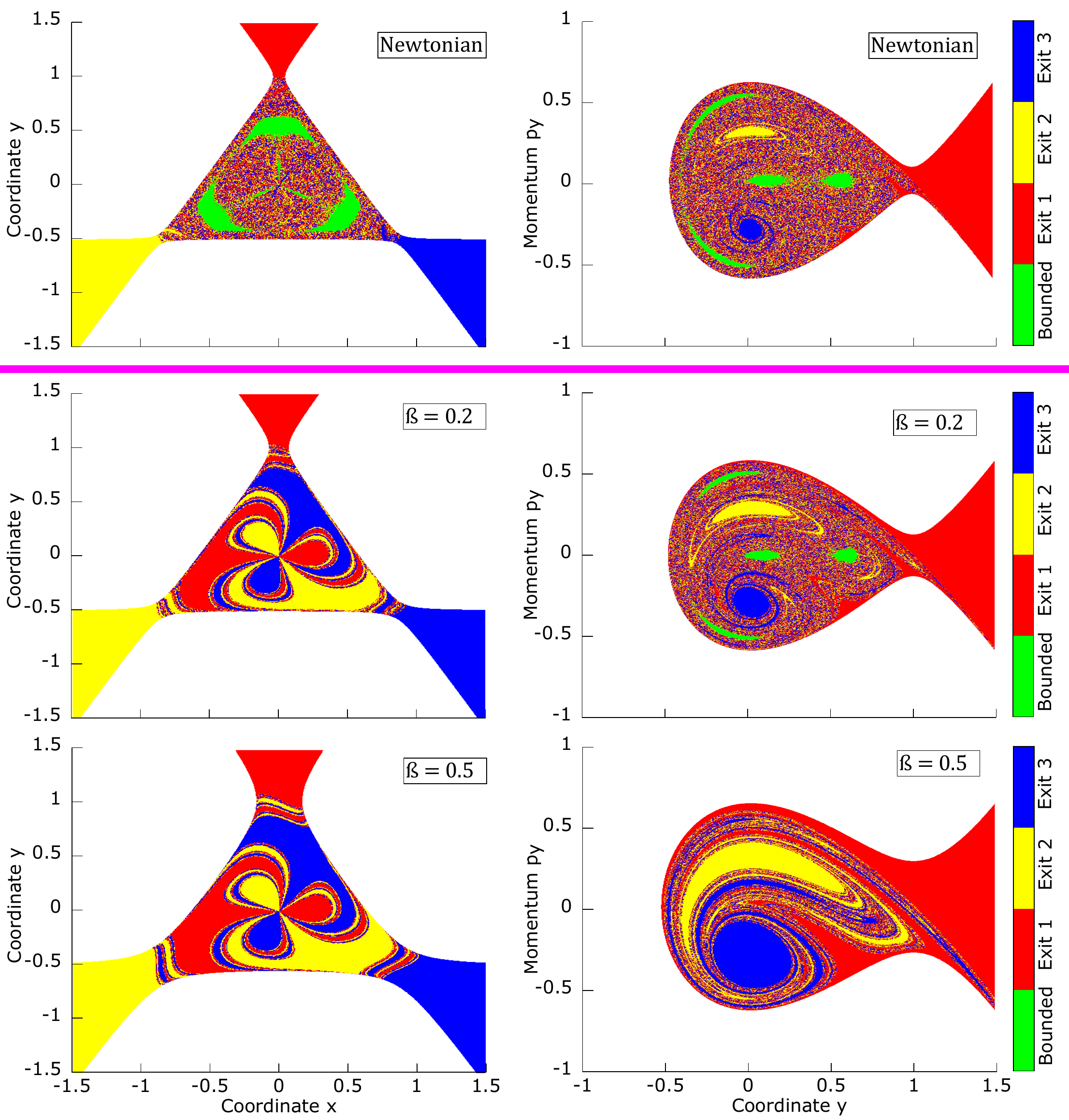}%
\caption{\label{lab:coord-phase-plane}\textbf{Exit basins of the H\'enon-Heiles Hamiltonian in the coordinates plane $(x,y)$ and the phase space $(y,p_y)$ for different values of $\beta$.}  The exit basins show the regions of initial conditions that lead to different escape outcomes or bounded trajectories in the H\'enon-Heiles system. The colors indicate the exit number (red (dark gray) for 1, yellow (white) for 2, blue (black) for 3) or bounded motion (green (light gray)). Initial conditions in the coordinates plane are chosen to maintain the symmetry of the problem, while in the phase space they satisfy $x(0) = 0$ and $p_x$ is obtained from the energy $E_N = 0.17$. The top row corresponds to the Newtonian case, while the central and bottom rows correspond to the relativistic cases  ($\beta=0.2$ and $\beta=0.5$, respectively).}
\end{figure}

\begin{figure}
\includegraphics[width=\textwidth]{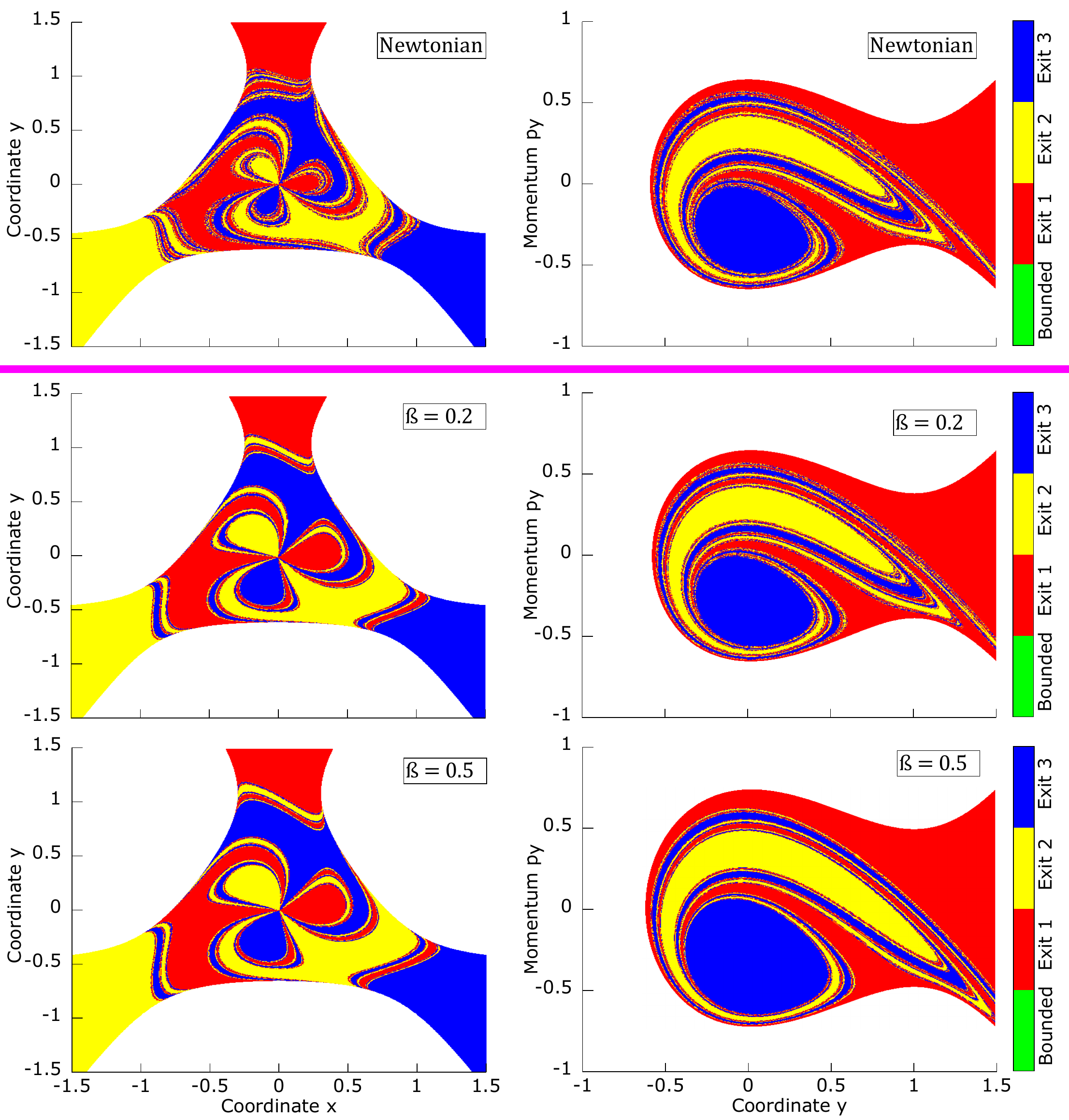}%
\caption{\label{lab:coord-phase-plane-025}\textbf{Exit basins of the H\'enon-Heiles Hamiltonian in the coordinates plane $(x,y)$ and the phase space $(y,p_y)$ for different values of $\beta$.}  The exit basins show the regions of initial conditions leading to different escape outcomes or bounded trajectories in the H\'{e}non-Heiles system. The colors indicate the exit number (red (dark gray) for 1, yellow (white) for 2, blue (black) for 3) or bounded motion (green (light gray)). Initial conditions in the coordinates plane are chosen to maintain the symmetry of the problem, while in the phase space they satisfy $x(0) = 0$ and $p_x$ is obtained from the energy $E_N = 0.25$. The top row corresponds to the Newtonian case, while the central and bottom rows correspond to the relativistic cases  ($\beta=0.2$ and $\beta=0.5$, respectively).}
\end{figure}

In this section, we investigate the behavior of the relativistic Hénon-Heiles system in both the physical space and the phase space. The physical space provides a spatial representation of the system's behavior, while the phase space provides a picture of its dynamics. We refer to Fig.~\ref{lab:coord-phase-plane} for illustrative examples.

We have used a grid of $1000 \times 1000$ initial conditions in order to know if the trajectories have escaped from the scattering region or if they remain bounded there after a maximum integration time of $1000$ time units. Afterwards,  we have plotted them on the coordinates plane ($x$,$y$) and the phase space ($y$, $p_y$). The trajectories have been colored based on their behavior: bounded trajectories have been colored green (light gray), while escaping trajectories have been colored blue (black) (Exit $1$), red (dark gray) (Exit $2$), or yellow (white) (Exit $3$), depending on the direction of escape, according to Fig.~\ref{Fig1}. We have used the Lyapunov orbit as a criterion for escape. The Lyapunov orbit is a periodic orbit that separates bounded and unbounded motions. In this kind of orbits, the particle is forced to escape to infinity and it never comes back when it crosses one of them in the outer direction \cite{AVS:2001}.

We have chosen the initial energy $E_N=0.17$ for the $(y,p_y)$ and $(x,y)$ planes, which corresponds to an initial velocity of $v_0=\sqrt{2E_N}=0.583095$ as commented in the model description. This value exceeds the escape energy $E_e=1/6$ of the Newtonian H\'enon-Heiles potential. We also set the relativistic parameter $\beta$ which measures the deviation from the Newtonian dynamics at high velocities to either $\beta=0.2$ or $\beta=0.5$.

Figure~\ref{lab:coord-phase-plane} shows the exit basins in both physical space and phase space. The basins in physical space are on the right, and the ones in phase space are on the left. The Newtonian energy is fixed at $E_N=0.17$ which corresponds to the open nonhyperbolic regime.  The upper panels correspond to the Newtonian case, where there are some regions of  bounded trajectories (green (light gray)), although most of the trajectories eventually escape (colored blue (black), red (dark gray) or yellow (white) according to the exit channel). In the middle and lower panels, we have denoted relativistic cases. For $\beta = 0.2$ in phase space, some bounded trajectories can still be observed.  The green (light gray) region shows trapping dynamics since the energy is smaller than the limit value and the relativistic value is not high enough to help the particles escape from the scattering region. As $\beta$  increases, the bounded regions become smaller and eventually vanish, indicating that all trajectories escape in the high relativistic regime. The phase space also reveals some additional bounded trajectories that are not visible in the coordinate plane due to different initial conditions. We observe that the relativistic factor $\beta$ plays a significant role in influencing the behavior of the trajectories.

When the Newtonian energy is in the open hyperbolic regime, we obtain the results of Fig.~\ref{lab:coord-phase-plane-025}, where all the trajectories escape. Figures  \ref{lab:coord-phase-plane} and  \ref{lab:coord-phase-plane-025} illustrate the influence of the Newtonian energy $E_N$ on the exit basins of the relativistic H\'{e}non-Heiles system in both the coordinate and the phase space. As $E_N$ increases from $0.17$ to $0.25$, the system undergoes a transition from the open nonhyperbolic regime to the hyperbolic regime, where all the periodic orbits become unstable and no KAM tori exist. This transition is reflected in the disappearance of the green (light gray) regions of bounded trajectories and the increase of the escape regions. Moreover, the exit basins become more sensitive to the relativistic factor $\beta$, which measures the deviation from the Newtonian dynamics at high velocities.

Higher values of $\beta$ tend to produce simpler and smoother structures due to the growing influence of the relativistic effects. This reduces the complexity and fractality of the exit basins and the sensitivity to the initial conditions. These results demonstrate the importance of considering the relativistic corrections when modeling open Hamiltonian systems with chaotic dynamics, especially for higher values of $E_N$, where the system is in hyperbolic regime and all trajectories escape.

\section{ESCAPES IN THE ENERGY PLANE $(y,H)$}\label{section_4}

\begin{figure}
\includegraphics[width=\textwidth]{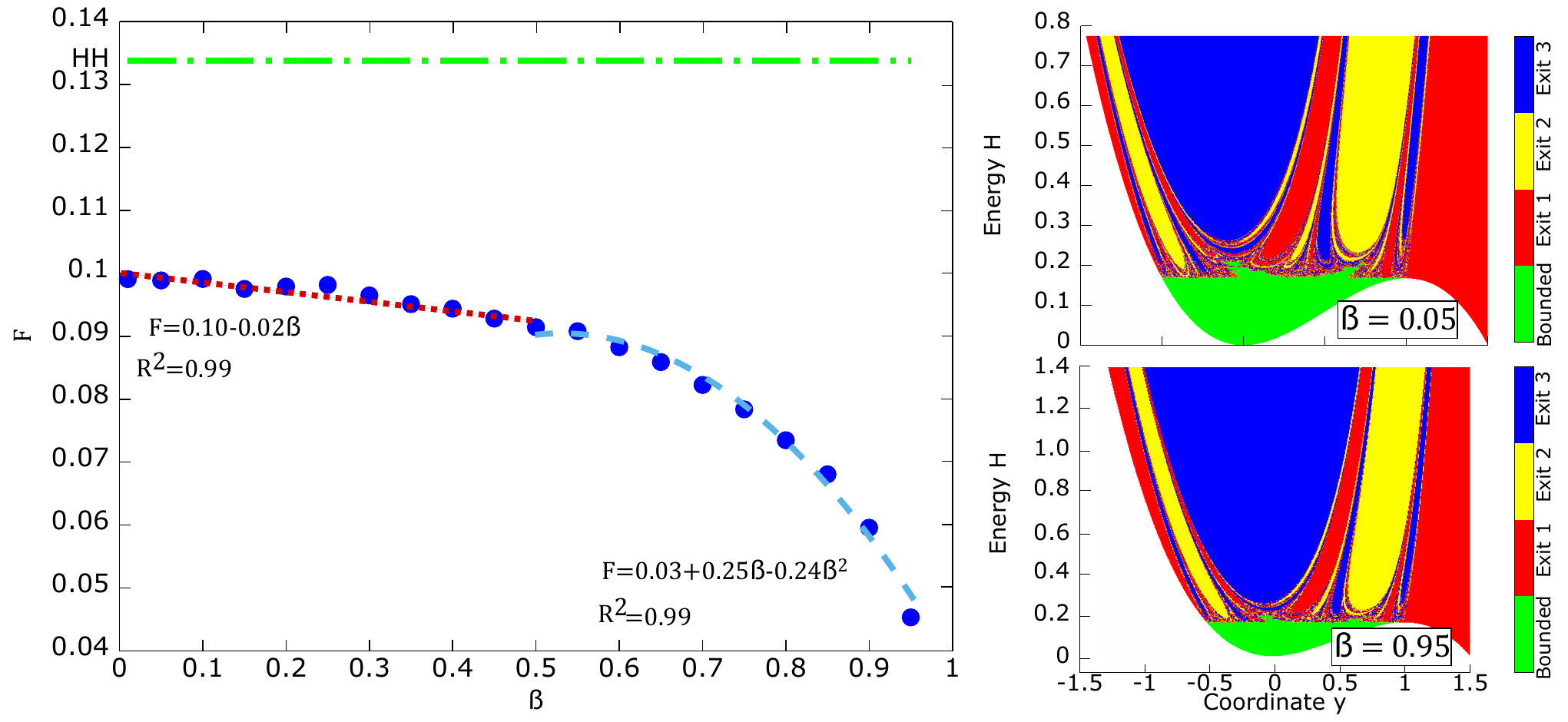}%
\caption{\label{energy_beta}  Left panel: $E_N=0.17$. Plot of bounded region size $F$ versus $\beta$ (in blue (black)). The linear fitting is denoted in red (dark gray), $F=0.1-0.02 \beta$, while the quadratic fit, $F=0.03+0.25 \beta -0.24 \beta^{2}$, has been plotted in light blue (light gray). Bounded region for the Newtonian H\'enon-Heiles is represented in green (light gray), where the particles are permanently trapped. Right panel: $(y,H)$ planes. Right above: $\beta=0.05$. Right below: $\beta=0.95$. In both cases, we can clearly see the effects of the high velocity for which the particles escape faster insofar $\beta$ is increasing.}
\end{figure}

Here, we study the fractal structures according to the value of the energy of the particle $E_N$.
Specifically, we compute the escapes for the $(y,H)$ plane, where $y$ is the vertical coordinate and $H$ is the energy of the relativistic case. We mean that $E=H = \sqrt{c^{4}+c^{2}p^{2}}+V(x,y)$, where $V(x,y)$ is the H\'{e}non-Heiles potential.
Here, we focus our attention in the regions of the plane $(y,H)$ in which we are mainly situated around the scattering region. Afterwards, we study how the particles are escaping from it for energy values between zero and the ones for which they are in the relativistic regime. We will also use this criterion in the remainder sections of this manuscript.
Right panels on Fig.~\ref{energy_beta} show the exit basins for $\beta=0.05$ and $\beta=0.95$, respectively, in the $(y,H)$ plane. The exit basins are color coded according to the behavior of the trajectories: green (light gray) for bounded trajectories, blue (black), red (dark gray), and yellow (white) for escaping trajectories through different exits. We note that $H$ depends on $\beta$, so it varies along the vertical axis when $\beta$ changes. This explains why the panel corresponding to the higher value of $\beta$ exhibits a significantly lower energy $H$ compared to the other panel.

We found that the size of the regular region, which is the region of bounded trajectories in the panels, remains relatively constant with respect to both $\beta$ and $E_N$, where $E_N$ is the Newtonian energy. Figure~\ref{energy_beta} shows the variation of the bounded region $F$ with $\beta$.
Each blue (black) dot in Fig.~\ref{energy_beta} represents the fraction of bounded trajectories out of the total number of possible trajectories in the chosen parameter region for a given relativistic energy $H$. Note that some combinations of initial values for the coordinate $y$ and the energy are forbidden. We observe that the fraction of bounded trajectories is always lower than the Newtonian value predicted by the H\'enon-Heiles model (green (light gray) line), regardless of the value of $\beta$. We can fit the fraction $F$ of bounded trajectories with high accuracy since $R^{2}=0.99$. The bounded region can be divided in $2$ different parts. The first part is the linear fit $F=0.1-0.02\beta$ which is denoted in red (dark gray). The second part fits as the quadratic fit $F=0.03+0.25\beta-0.24\beta^2$ which is shown in light blue (light gray). The change of the behavior takes place at $\beta \simeq 0.5$ in which the KAM islands are completely destroyed and the particles escape from the scattering region very fast. This is the reason for which the fitting between the fraction of the particles trapped versus $\beta$ changes from a linear to a quadratic scaling. As it was shown in previous works \cite{JD relativistic}, the destruction of the KAM islands involves a bifurcation in the dynamics of the system, from an algebraic decay law to an exponential decay law of the particles. That has strong implications in the determination of global properties of the system such as the average escape time of the particles. In our case, this last result is relevant in the sense that we have a scaling law between the fraction of the remaining particles versus $\beta$. Therefore, we can predict the evolution of the escaping dynamics as a function of the velocity of the particles. Besides, this behavior must be general since the H\'{e}non-Heiles system is a paradigmatic system in chaotic scattering. Therefore, we expect that this scaling law and the ones we show in the next section are from a general nature from a qualitative point of view.

\begin{figure}
\includegraphics[width=\textwidth]{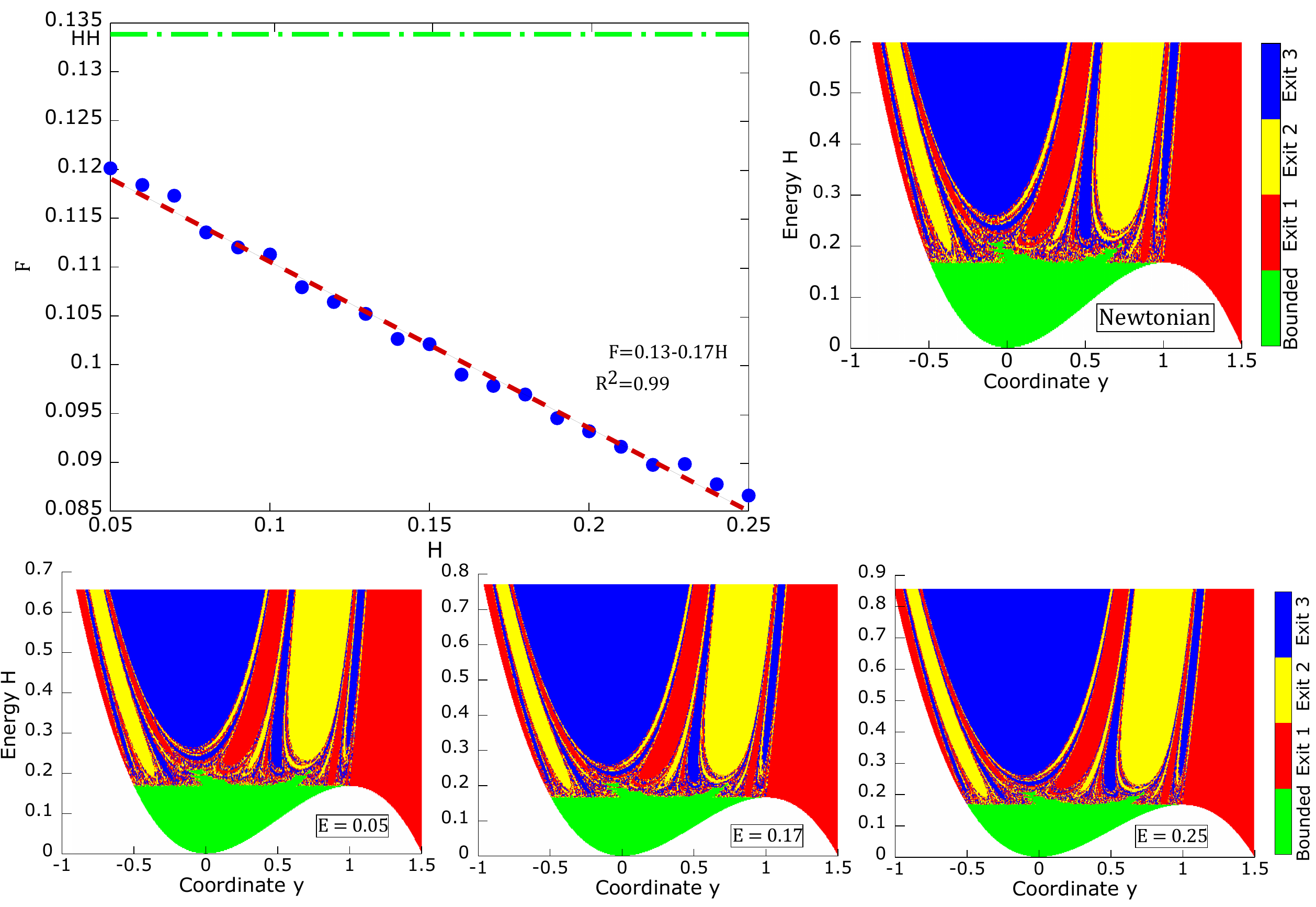}%
\caption{\label{energy_h} Plot of the bounded region $F$ (assuming a total size of $1$) versus the relativistic energy $H$ for $\beta = 0.2$ (in blue (black)). The  linear fit,  $F=0.13-0.17 H$, is plotted in red (dark gray). The picture on the right side (up) represents the situation for the Newtonian H\'{e}non-Heiles system. In green (light gray), we plot the bounded trajectories. The $(E_N, y)$ planes below show different situations of the relativistic regime for $\beta=0.2$ and energies $E_N=0.16$, $E_N=0.2$ and $E_{N}=0.25$ from left to right, respectively. As in Fig.~\ref{energy_beta}, the horizontal dash line in green (light gray) color denotes the bounded region $F$ for the Newtonian H\'{e}non-Heiles system in which the particles are permanently trapped.}
\end{figure}

Figure~\ref{energy_h} presents a comparison between the Newtonian model in the ($y$,$H$) plane (top) and three relativistic situations with to $E_N=0.05$, $E_N=0.16$, and $E_N=0.25$, respectively (below from left to right) in which $\beta=0.2$. The plots show that for $E_N=0.16$, which is just below the escape energy of $E_e=1/6$, there are still some regions of bounded trajectories (green (light gray)), while for $E=0.25$, which is above the escape energy, all trajectories have escaped (denoted in blue (black), red (dark gray), and yellow (white) colors, respectively). Besides, in Fig.~\ref{energy_h}, the variation of the size of the regular region, which is the region of bounded trajectories in the plates, as a function of the relativistic energy $H$ is shown. The bounded region is plotted in blue (black). We can also find a fitting between the fraction of trapped particles and $H$ with a very good precision since $R^{2}=0.99$. A linear fitting is found which is relevant since the fraction of remaining particles in the scattering region decreases linearly with the energy according to the law $F=0.13-0.17H$ with $R^{2}=0.99$. The green (light gray) line represents the Newtonian value for the H\'enon -Heiles potential.
As in the previous figure, a scaling law is obtained for which we can predict the evolution of the remaining particles in the region before escaping as a function of the relativistic energy $H$.

\section{ESCAPES IN THE $\beta$ PLANE $(y, \beta)$}\label{section_5}

In order to complete our study on different fractal structures in relativistic chaotic scattering, we include the computation of the basins by varying the value of the relativistic parameter $\beta$ and the variable $y$.
For this purpose, we investigate the dynamics of the $(y,\beta)$ plane, where $y$ is the vertical coordinate and $\beta$ is the relativistic factor. We varied $\beta$ from $0$ to $1$ for different values of the Newtonian energy $E_{N}$, which then caused a change of the initial velocity $v$ and the relativistic energy $H$.

\begin{figure}
\includegraphics[width=0.75\textwidth]{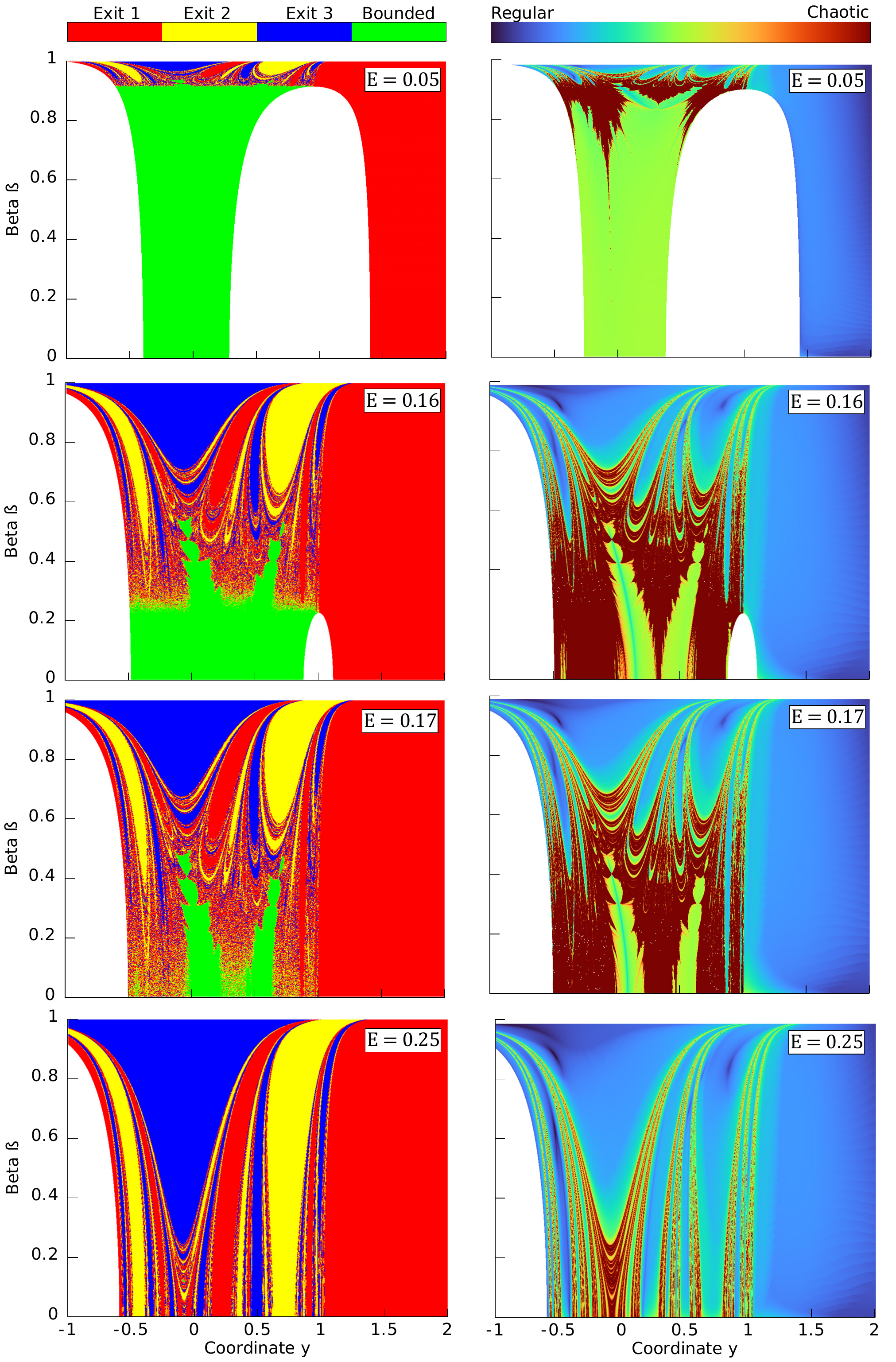}%
\caption{\label{beta_plane} Plot of the $(y,\beta)$ plane for $E_N = 0.05, 0.16, 0.17, 0.25$. Left panes: colors indicate the exit number (red (dark gray) $1$, yellow (white) $2$ or blue (black) $3$) or bounded motion (green (light gray)), as in the previous figures. Right panels: colors indicate the value of the chaos indicator $OFLI2$. Regular regions (fast escape) (blue (light gray)), chaotic trajectories (red (dark gray)), periodic trajectories (green (white)), quasiperiodic (yellowish (pale gray)). The color code is as described in the captions of the previous figures. Notice that for $E_N=0.25$ there are escapes for all values of $\beta$ which show that KAM islands are destroyed.}
\end{figure}

As $E_{N}$ increases, the structures in the $(y,\beta)$ plane undergo significant changes, as illustrated in the left panels of Fig.~\ref{beta_plane}. We use the same color-coding scheme as in previous sections to represent the different types of trajectories: green (light gray) for bounded trajectories and blue (black), red (dark gray), and yellow (white) for escaping trajectories through different exits. For low values of $E_N$, there is a large region of bounded trajectories (green (light gray)), while for high values of $E_N$, most trajectories escape (blue (black), red (dark gray), or yellow (white)). The right panels of Fig.~\ref{beta_plane} show the $OFLI2$ chaos indicator \cite{BBS} for each trajectory visualized using the Turbo colormap. This colormap is a perceptually uniform and colorblind-friendly colormap that was introduced by Google. It ensures a smooth transition of colors that avoids abrupt changes in hue and lightness. The chaos indicator serves as a valuable tool to distinguish between regular trajectories (the intensity of the blue (black) color indicates that the trajectory is more regular) and chaotic trajectories (denoted by the intensity of the red color (dark gray) in those regions). Escape trajectories are regular, since they asymptotically approach infinity and their dynamics do not vary. Periodic or quasiperiodic trajectories appear in green (light gray) or yellowish colors (pale gray).

\begin{figure}
\includegraphics[width=0.7\textwidth]{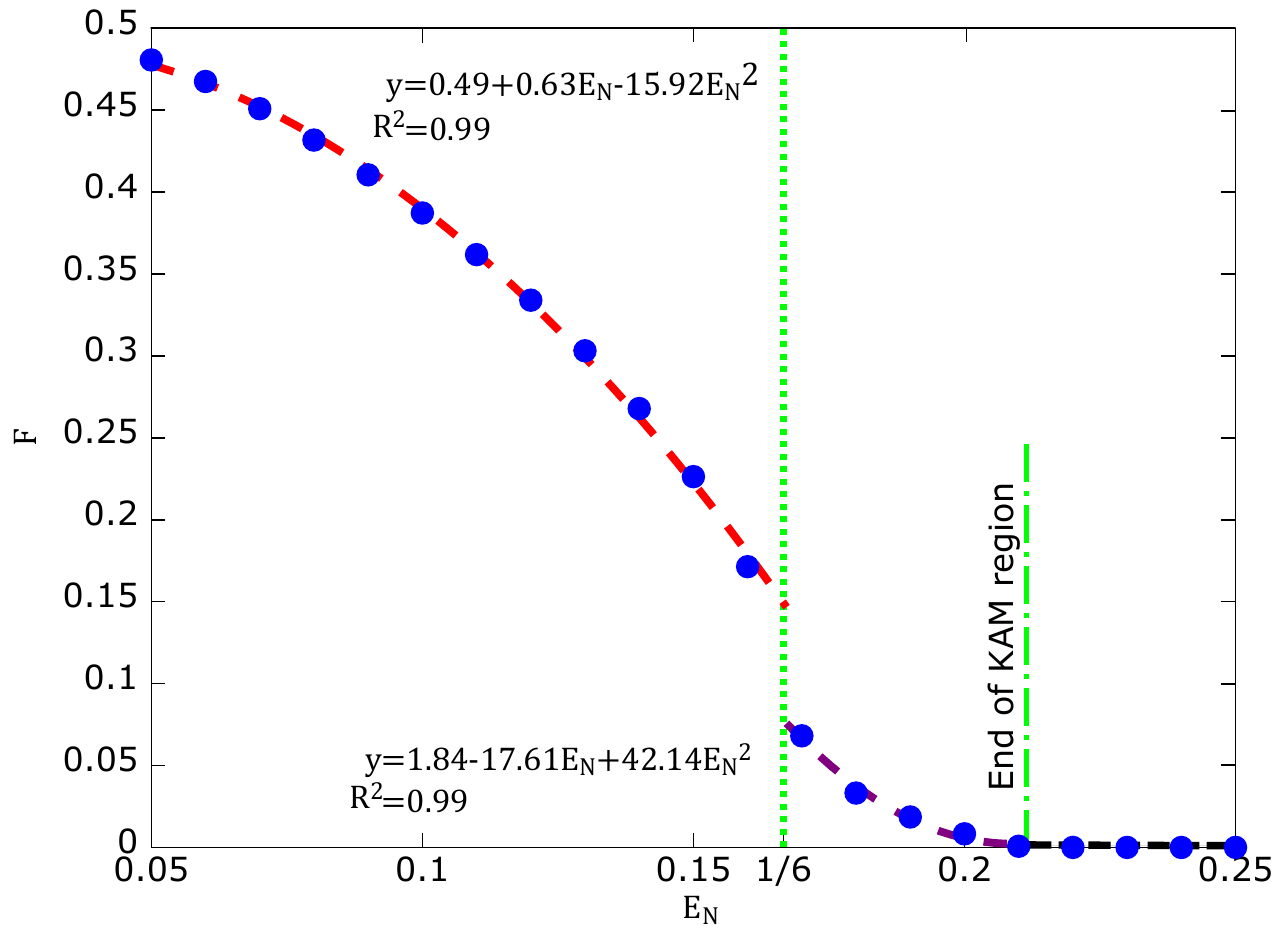}%
\caption{\label{beta_2fit} Quadratic fits for energy values from $E_N=0$ to the end of KAM region. The escape Newtonian energy, $E_e=1/6$, is denoted by a dashed green (pale gray) vertical line. The quadratic fit $F=0.49+0.63E_{N}-15.92 E_{N}^{2}$ changes very suddenly to $F=1.84-17.61E_{N}+42.14 E_{N}^{2}$ when the particles have enough energy to escape ($E_{N}>E_{e}=1/6$). The jump at $E_{N}=1/6$ is due to the trajectories start to escape from the scattering region and, therefore, the system becomes open presenting three different channels for which the particles can escape. The second dashed green (light gray) vertical line (located at $E_{N} \cong 0.21$) denotes the end of the KAM islands. There, all trajectories are escaping since it is the beginning of the hyperbolic regime and, therefore, $F\rightarrow 0$.}
\end{figure}

Now, we are interested in the relationship between the fraction of the trapped trajectories $F$ and the Newtonian energy $E_{N}$ in order to compare with the previous section in which we computed $F$ versus the relativistic energy $H$. Here, the size of the regular region, which is the region of bounded trajectories in the panels, is shown in Fig.~\ref{beta_2fit} as a function of $E_N$. Since the size does not correspond to a single curve, we fit two quadratic curves for different ranges of $E_N$ with a high accuracy in both cases ($R^{2}=0.99$). The first curve (red (dark gray)) fits the data for $E_N<0.16$, which is below the escape energy of $E_e = 1/6$. This law is $F=0.49+0.63E_N-15.92 E_N^2$. The second curve (magenta), $F=1.84-17.61E_N+42.14E_N^2$, fits the data for $0.16<E_N<0.25$, which is above the escape energy but still has some regions of bounded trajectories due to KAM islands. The curves show that the size of the regular region decreases with increasing $E_N$. For $E_N>0.25$, there are no KAM islands and only small zones of stable motion remain, which can be modeled by a horizontal line (black). The sudden transition which takes place at $E_{N}=1/6$ in which there is a drastic change in $F$ is due to the dynamical consequences for the particles when they reach the escape energy $E_{e}$ (denoted by a dashed vertical green (light gray) line). Afterwards, for that value of the Newtonian energy, $E_{N}>E_{e}=1/6$,  the particles are starting to escape from the scattering region and the fraction of the remaining particles decreases faster than when $E_{N}<E_{e}=1/6$. Finally, when the energy $E_{N}\approx 0.25$ (denoted by another dash green (pale gray) vertical line) the regime is hyperbolic and all particles escape. Therefore, here, $F\rightarrow 0$ as all KAM islands disappear.
We can clearly observe the difference between this figure and Fig.~\ref{energy_h} where a linear fitting between $F$ and $H$ was found and where we can see that $F\rightarrow 0$ as $H\rightarrow 0.25$ as in the Newtonian case.

Our examination of the behavior of the $(y,\beta)$ plane for different values of $E_N$, reveals that the structures change significantly as $E_N$ increases. The size of the regular region is shown to decrease with increasing $E_N$ and can be modeled by two quadratic curves for different ranges of $E_N$. Once the KAM region ends, there are only small zones of stable motion and therefore the fraction of bounded trajectories approaches zero. These findings contribute to our understanding of the dynamics of the $(y,\beta)$ plane providing a way to predict how the particles are escaping from the scattering region as a function of the relativistic parameter $\beta$.


\section{Conclusions and discussion}\label{sec_conclusions}

The analyzed structures in relativistic chaotic scattering model that we have investigated have a very rich fractal character and the dynamics of the particles is quite complex.
The numerical analysis of the different structures provides relevant information of the dynamics of the system and the associate phase space topology. For this purpose, we have used the relativistic version of the H\'{e}non-Heiles system as a paradigmatic model in chaotic scattering. First, we have shown that the physical space and the phase space have a very rich fractal structure where the increasing of the energy clearly shows the particles to escape faster from the scattering region as compared with the Newtonian case. The energy of the system plays a crucial role in the escaping dynamics and the numerical analysis of the $(y, H)$ plane reveals a scaling law between the fraction of particles remaining in the scattering region $F$. This scaling law is a quadratic function between the trapped particles $F$ and the relativistic parameter $\beta$. A similar study of these structures in the $\beta$ plane reveals a linear scaling law between the fraction of particles in the scattering region and the relativistic energy $H$. This is a quite relevant result since the escapes and the energy, in the Newtonian case, are related in a linear way, which is important for the prediction of the evolution of the system. However, this scaling law becomes quadratic when we consider the Newtonian energy $E_N$. In this last situation, we can fit the curve in $2$ regions: when the energy is below the escape energy and when the energy is higher than this value becoming a quadratic fitting in both cases. One possible and useful application of the results of this paper is in the study of charged particles which are trapped under the effects of a dipole magnetic field. In this physical situation, the study of these kind of structures provides fundamental insights to characterize how the particles are escaping or are trapped by the action of the magnetosphere. In this last situation, these structures can also provide information on the chaotic behaviors of protons and electrons which are moving in bounded regions around the Earth \cite{DIPOLE:2022}. It is crucial to note that in such specific problems, an appropriate system of units
(typically, the International System or the centimeter-gram-second system) should be employed to
accurately handle real physical situations. Finally, we expect that the results presented here can be useful for a better understanding of the chaotic scattering phenomena in both Newtonian and relativistic regimes which have implications in several field in physics.

\section{Acknowledgment}

This work has been supported by the Spanish State Research Agency (AEI) and the European Regional Development Fund (ERDF, EU) under Project No.~PID2019-105554GB-I00 (MCIN/AEI/10.13039/501100011033).


\end{document}